\documentclass[9pt,conference]{IEEEtran}
\IEEEoverridecommandlockouts
\usepackage{cite}
\usepackage{amsmath,amssymb,amsfonts}
\usepackage{algorithmic}
\usepackage{graphicx}
\usepackage{textcomp}
\usepackage{url}
\usepackage[dvipsnames]{xcolor}
\usepackage{hyperref}
\usepackage{multirow}
\usepackage{caption}
\usepackage{etoolbox}
\usepackage{makecell}
\usepackage{enumitem}
\setlist{nolistsep}

\def\BibTeX{{\rm B\kern-.05em{\sc i\kern-.025em b}\kern-.08em
    T\kern-.1667em\lower.7ex\hbox{E}\kern-.125emX}}

\begin{document}

\newcommand{\insertfig}{
\centering{\includegraphics[width=\textwidth]{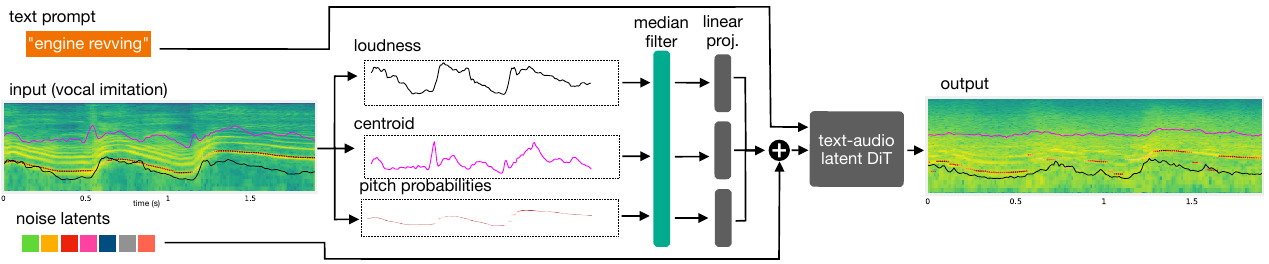}}
\captionof*{figure}{Fig 1. Overview of Sketch2Sound. We extract three control signals from any input sonic imitation: loudness, spectral centroid (i.e.,~brightness) and pitch probabilities. We encode the signals and add them to the latents used as input to a DiT text-to-sound generation system. \textbf{We \textit{strongly} encourage the reader to listen to this example and many more at: \textcolor{blue}{\href{https://hugofloresgarcia.art/sketch2sound}{https://hugofloresgarcia.art/sketch2sound}}}.
}
\label{fig:overview}
}
\makeatletter
\apptocmd{\@maketitle}{\centering\insertfig}{}{}
\makeatother

\title{Sketch2Sound: Controllable Audio Generation via Time-Varying Signals and Sonic Imitations\\

}

\author{\IEEEauthorblockN{Hugo Flores García$^{io}$, Oriol Nieto$^{i}$, Justin Salamon$^{i}$,  Bryan Pardo$^{o}$, Prem Seetharaman$^{i}$}
\textit{$^{i}$Adobe Research, $^{o}$Northwestern University}\\
hugofloresgarcia@u.northwestern.edu

}

\maketitle


\begin{abstract}

We present Sketch2Sound, a generative audio model capable of creating high-quality sounds from a set of interpretable time-varying control signals: loudness, brightness, and pitch, as well as text prompts. Sketch2Sound can synthesize arbitrary sounds from sonic imitations (i.e.,~a vocal imitation or a reference sound-shape). Sketch2Sound can be implemented on top of any text-to-audio latent diffusion transformer (DiT), and requires only 40k steps of fine-tuning and a single linear layer per control, making it more lightweight than existing methods like ControlNet. To synthesize from sketchlike sonic imitations, we propose applying random median filters to the control signals during training, allowing Sketch2Sound to be prompted using controls with flexible levels of temporal specificity. We show that Sketch2Sound can synthesize sounds that follow the gist of input controls from a vocal imitation while retaining the adherence to an input text prompt and audio quality compared to a text-only baseline. Sketch2Sound allows sound artists to create sounds with the semantic flexibility of text prompts and the expressivity and precision of a sonic gesture or vocal imitation. Sound examples are available at \textcolor{magenta}{\url{https://hugofloresgarcia.art/sketch2sound/}}.
\end{abstract}


%
\section{Introduction}


Sound design is the craft of storytelling through sonic composition. Within sound design, Foley sound is a technique where special sound effects are designed and performed in sync to a film during post-production \cite{ament2021foley}. These sound scenes are typically performed by a Foley artist on a stage equipped with abundant sound instruments and other soundmaking materials\footnote{Example of a Foley artist performing a scene: \href{youtu.be/WFVLWo5B81w}{youtu.be/WFVLWo5B81w}}. Foley sound is a skilled and gestural performance art: performing a sound scene with sound-making objects and instruments (instead of arranging pre-recorded samples post hoc) allows sound artists to create fluent and temporally aligned sounds with a ``human'' (i.e.,~gestural) touch. 
Adding this gestural touch to the resulting sound composition often results in a sonic product of great aesthetic and production value. 


Recent research in generative modeling for sound has paved the way for text-to-sound systems\cite{evans2024longform, yang2023diffsound, liu2023audioldm}, where a user can create sound samples from text descriptions of a sound (e.g.,~``explosion''). While the text-to-sound paradigm can help a sound designer find sounds more quickly (and, perhaps in the future, with a higher degree of specificity), a sound designer still has to painstakingly modify the temporal characteristics of the generated sound so that they can be in sync with the visuals in the editing timeline.
This is in opposition to the natural way that Foley artists gesturally create sound effects by physically performing with physical soundmaking objects. 


To overcome the drawbacks of a purely text-to-audio interaction, several works in the music domain sought to condition generative models on audio \cite{floresgarcia2023vampnet}, parallel instrument stems \cite{parker2024stemgen}, melody \cite{copet2023simple}, sound event timestamps and frequency \cite{xie2024picoaudio}, or multiple structural control signals like song structure and dynamics \cite{novack2024ditto}.  Notably, \cite{devis2023continuous} condition an audio VAE on control signals such as brightness and loudness, though their experiments are limited to models trained on narrow sound distributions (e.g.,~violin, darbouka, speech) and not a multi-distribution text-to-audio model. For speech, \cite{morrison2024finegrainedinterpretableneuralspeech} proposes a fully interpretable and disentangled representation for speech generation and editing, which allows for fine-grained control over the pitch, loudness, and phonetic pronunciation of speech.



\textit{The human voice is a gestural sonic instrument} \cite{wishart1988composition}: it allows us to realize sounds without having to perform any symbolic abstraction (i.e.,~putting a sound into words) beforehand.
When humans communicate audio concepts to other people, they typically combine  language and vocal imitation \cite{lemaitre2014effectiveness, lemaitre2016vocal, lemaitre2011vocalimitations}, 
and recent work has shown its utility for query-by-example search of audio databases \cite{zhang2020vroom, blancas2014sound, zhang2016imisound }.
This is a more natural method than describing the evolution of pitch, timing, and timbre via pure text descriptions \cite{lemaitre2014effectiveness}, and voice-driven sound synthesis interactions have been of interest long before modern generative modelling for their embodied capabilities\cite{dellemonache2018embodiedsounddesign, hazan2005billaboop, janer2008extending, piccolo2017non, fasciani2013mapping, baldan2016sketching}. 


\textcolor{black}{\textbf{We propose Sketch2Sound: a text-to-audio model that can create high-quality sounds from sonic imitation prompts by following interpretable, fine-grained time-varying control signals that can be easily extracted from any audio signal at different levels of temporal detail:  loudness, brightness (\textit{spectral centroid}) and pitch}.}
We expand upon previous work \cite{chung2024tfoley} by developing a method capable of following the loudness, brightness \textit{and} pitch of a vocal imitation, with the option to drop any of the three controls. Additionally, we propose a technique that varies the temporal detail of the control signals used during training by applying median filters of different window sizes to the control signals before using them as input. This allows sound artists to specify the degree of temporal precision to which a generative model should follow the specified control signals, which improves sound quality in sounds that may be too hard to perfectly imitate with one's voice.

\textbf{This method is not limited to just vocal imitation}: any kind of sonic imitation can be used to drive our proposed generative model --  we place the focus on vocal imitation due to people's innate ability to imitate sounds with our voices. 
Vocal imitations can always be augmented through other sonic gestures like clapping, tapping, playing instruments, etc. 
Sketch2Sound can be added to any existing latent diffusion transformer (DiT) sound generation model with as little as 40k fine-tuning steps. Unlike ControlNet methods \cite{wu2023music, zhao2023uni}  that require an extra trainable copy of the entire neural network encoder, Sketch2Sound requires only a single linear layer per control.

Our experiments show that Sketch2Sound can generate sounds that closely follow the input control signals (loudness, spectral centroid, and pitch/periodicity) from a vocal imitation while still achieving a high degree of adherence to a text prompt and an audio quality comparable to the text-only pre-trained model. 
We show that our median filtering technique leads to improved audio quality and text adherence when generating sounds from vocal imitations. 
We also show that, during inference, a user can arbitrarily specify a degree of temporal detail by choosing a median filter size, allowing them to navigate the trade-off between strict adherence to the vocal imitations and audio quality + text adherence.  

To the best of our knowledge, this is the first sound generation model capable of following vocal imitations \textit{and} text prompts by conditioning on a set of holistic control signals suitable for generating sound objects with fine-grained, gestural control of pitch, loudness, and brightness. 
We believe Sketch2Sound will give sound artists a more expressive, controllable, and gestural interaction for generating sound-objects than existing text-to-audio and other conditional sound generation systems. \textbf{\textcolor{black}{We highly encourage the reader to listen to our expansive set of audio examples demonstrating Sketch2Sound. }}
\footnote{\href{https://hugofloresgarcia.art/sketch2sound}{https://hugofloresgarcia.art/sketch2sound}}

\begin{table*}[htbp]
\caption{Control Signal Evaluation and Sketch Type Ablation. \textbf{Control Signals}: Introducing more control signals improves control adherence while trading off text adherence and audio quality. \textbf{Sketch Types:} Using median filters to create sketch curves achieves the best trade-off between text adherence/audio quality and control adherence, compared to a no filtering baseline and a low pass filtering approach.}
\begin{center}
\setlength{\tabcolsep}{4pt} 
\renewcommand{\arraystretch}{1.5} 
\begin{tabular}{|c|c|c|c|c|c|c|c|c|c|}
\hline
\textbf{} & \textbf{Control Signal} & \textbf{\makecell{\\Sketch \\Type}} & \textbf{\makecell{Text \\Adherence $\uparrow$}} & \multicolumn{4}{|c|}{\textbf{Control Adherence (Error) $\downarrow$}} & \multicolumn{2}{|c|}{\textbf{\makecell{Audio Quality\\ (FAD) $\downarrow$}}} \\
\cline{5-10}
 & & & \textbf{CLAP Score} & \textbf{RMS (dB)} & \textbf{Centroid (st)} & \textbf{Pitch (st)} & \textbf{Chroma (st)} & \textbf{VGGish} & \textbf{CLAP} \\
\hline
\multirow{4}{*}{\makecell{\textit{control} \\ \textit{signals}}} 
& text-only        & median (sz 10)                                & \b{0.273}  & 13.41 & 10.34 & 13.91 & 2.96 & 2.57 & 0.27 \\
\cline{2-10}
& loudness (ldns)  &  median (sz 10)                               & 0.230  & 3.88  & 10.37 & 12.45 & 2.96 & 2.69 & 0.296 \\
\cline{2-10}
& ldns+centroid &  median (sz 10)                                 & 0.219  & 3.60  & 4.39  & 11.17 & 2.87 & 2.67 & 0.306 \\
\cline{2-10}
& \textbf{ldns+centroid+pitch (ours)} &  median (sz 10)           & 0.211  & 3.60  & 4.43  & 1.49  & 0.48 & 2.51 & 0.312 \\
\hline
\hline
\multirow{2}{*}{\makecell{\textit{sketch} \\ \textit{types}}} 

& ldns+centroid+pitch & low pass                                  & 0.166  & 2.19  & 3.33  & 0.44  & 0.23 & 3.30 & 0.363 \\
\cline{2-10}
& ldns+centroid+pitch & no filters                                & 0.152  & 1.87  & 3.21  & 0.45  & 0.21 & 3.53 & 0.379 \\
\hline
\end{tabular}
\label{tab:combined_table}
\end{center}
\end{table*}

\section{Method}
We propose a method for conditioning an audio latent diffusion model on a set of interpretable, time-varying control signals that are suitable tasks creating variations of sounds and generating new sounds expressively via text-prompted sonic imitations.

\subsection{Time-varying control signals for sound objects}

We use the following control signals as conditioning for Sketch2Sound: 

\begin{itemize}
    \item \textbf{Loudness:} 
    We extract the per-frame loudness of an audio signal by performing an A-weighted sum across the frequency bins in a magnitude spectrogram \cite{morrison2024finegrainedinterpretableneuralspeech} and taking the RMS of the result.
    
    \item \textbf{Pitch and Periodicity:} 
    We use the raw pitch probabilities of the CREPE \cite{kim2018crepeconvolutionalrepresentationpitch, morrison2022torchcrepe} (``tiny'' variant) pitch estimation model. To avoid leaking timbral information in this signal, we zero out all probabilities below 0.1 in the pitch probability matrix. 
    
    \item \textbf{Spectral Centroid} 
    is defined as the center of mass of the frequency spectrum for a given audio frame. Frames with a higher centroid will be perceived as having a brighter timbre.  To preprocess the centroid, we convert the signal from linear frequency space (i.e.,~Hz)  to a continuous MIDI-like representation, scaled to roughly a $(0, 1)$ range by dividing the input signal by $127$ (note G9, roughly 12.5kHz), which we found to stabilize the first steps of training. 
\end{itemize}

Other momentary time-varying control signals may be used as well.  

\subsection{Conditioning a latent audio DiT on time-varying control signals}
Refer to Figure 1 for a visual overview of our approach. We use a large pre-trained text-to-sound latent diffusion transformer (DiT), similar to the one described in \cite{evans2024longform, evans2024fast} (text-conditioned only, no timing conditioning) and adapt it to generate sounds conditioned on the time-varying control signals mentioned above. 
The latent diffusion model for text-to-sound generation has two parts: first, a variational autoencoder (VAE) compresses 48kHz mono audio to a sequence of continuous vectors of dim 64 at a rate of 40Hz. Then, a transformer model is trained to generate new sequences of latents, which can be decoded into audio using the VAE decoder. 
This text-to-audio DiT was pre-trained on a large mix of proprietary, licensed sound effect datasets and publicly available CC-licensed general audio datasets. Once the model is pre-trained, we fine-tune it for 40k steps and adapt it to handle our time-varying control signals.

Because the time-varying control signals can be easily and efficiently extracted from any audio signal on the fly, we can fine-tune the pre-trained text-to-audio model in a self-supervised manner: Given any input audio signal, we extract the three control signals (loudness, centroid, pitch) from the audio signal and use them as conditioning for the model during fine-tuning. The model is then fine-tuned using the same recipe used during training: learning the reverse diffusion process from a set of noisy latents with text conditioning, along with our proposed control conditioning. 

To align the time-varying control signals with the latents from our text-to-sound DiT, the control signals must be extracted at the same frame rate as the audio VAE latents or interpolated to this frame rate. This allows us to perform a simple conditioning method: condition a latent diffusion model $\epsilon_{\theta}$ by simply adding a linear projection layer from our control signals to the noisy latents used as input to the diffusion model. Since these time-varying control signals are highly localized to their given time frame, a simple linear layer suffices to incorporate each time-varying signal as conditioning to the model.

Given the noisy latent vector sequence $\mathbf{z} \in  \mathbb{R}^{D \times N}$ that is used as input to a latent diffusion model $\epsilon_{\theta}$ with embedding dimension $D$ and sequence length $N$, we introduce our time-varying conditioning signal $\mathbf{c}_{ctrl} \in \mathbb{R}^{K \times N}$ with dimension $K$ and sequence length $N$ to the latent $\mathbf{z}$ by applying a trainable linear projection layer $p_\theta(\mathbf{c}_{ctrl}) \in \mathbb{R}^{D \times N}$ to the input conditioning, and adding the result directly to the latents used as input to the diffusion model:
$ \mathbf{z}_{ctrl} = p_\theta(\mathbf{c_{ctrl}}) + \mathbf{z} $.
We can repeat this process for any number of time-varying control signals that we'd like to condition our latent diffusion model. 

During fine-tuning, the loss configuration for the model does not change from the one used in original training; that is, we do not apply any reconstruction losses for the control curves themselves, removing the need to pass through the VAE decoder during fine-tuning. Despite not measuring loss on the control signals, providing them as input during fine-tuning is sufficient for the model to condition generation on them. As a result, these signals become useful for control at inference time. 
To ensure we can generate without requiring all the control signals, we perform dropout during fine tuning on the control signals by zero-ing out the control embeddings $p_\theta(\mathbf{c_{ctrl}})$ before they are added to the diffusion model's latents $\mathbf{z}$. We drop each control signal (as well as the text conditioning) individually with a 20\% probability, with an added 20\% probability of dropping out all signals together. Likewise, these signals can be dropped out at inference if a user wishes to control only one or two controls while omitting others.

At inference time, we follow the two-conditioning classifier-free guidance setup described in \cite{brooks2023instructpix2pixlearningfollowimage}. We use guidance scales $s_{ctrl}$ and $s_{text}$, which can be used to trade off the guidance strength for the control signal conditioning and text conditioning, respectively. We find that using a single guidance scale for all three control signals together ($s_{ctrl}$) is a sufficient approach, but future work may explore the effect of applying guidance strengths to each time-varying control independently of each other.   
Anecdotally, we find that setting $s_{text}$ to a value of $5$ and $s_{ctrl}$ to $1$ achieves results with good text adherence while following the contour provided in the sonic imitation controls. 


\subsection{Creating sketchlike controls via control-rate filtering}
Even though the human voice is remarkable at imitating sounds, there will still be a mismatch between the control signals of the target sound and the control signals of the vocal imitation.  
To overcome this issue, we propose a technique to make the controls sketchlike by applying random median filters to the control signals at different window sizes (1-25 control frames) before they are used as input to the model.
This filtering technique can help mitigate the mismatch in temporal specificity between the vocal imitation and target sound and help the model produce higher-quality sounds from sketchlike vocal imitations. 
Our experiments show that during inference, a sound artist is free to adjust the control rate of their input control signals, giving them an interpretable control over the trade-off between text-prompt adherence and fine temporal precision.

\section{Experimental Design}  \label{sec:exp-design}

Our experiments evaluate Sketch2Sound's ability to synthesize high-quality sounds from vocal imitations.  
Except for the text-only baseline (which doesn't need fine-tuning), we fine-tune every model for 40k  steps with the same configuration used for training. For all fine-tunings, we use the text-only baseline as the starting checkpoint.  

The main dataset used for evaluation is VimSketch \cite{kim2019vimsketch}, which consists of approximately 12k vocal imitations, each with a text description and reference sound. For each model variant, we generate 10k examples with durations of up to 5 seconds using the vocal imitation and text description as conditioning. 
We evaluate Sketch2Sound along the following characteristics:



 \textbf{Audio Quality}: To measure a model's ability to synthesize high-quality sound effects, we compute the Frechét Audio Distance (FAD) \cite{kilgour2018fad} using a proprietary dataset of 40k high-quality sound effects as the reference set, and 10k sounds generated from vocal imitations from the VimSketch dataset as the evaluation set, as suggested by \cite{fadtk}. We report the FAD for VGGish \cite{vggish} and LAION-CLAP \cite{laionclap2023} embeddings.
\textbf{Text Adherence:} We measure how well our generated audio adheres to the target text prompt by computing the CLAP embedding cosine similarity \cite{laionclap2023} between audio generated from a sonic imitation and the target text prompt for every example.  

\textbf{Control Signal Adherence:} Finally, to measure the adherence of the generated audio to the vocal imitations, we measure the error (L1) between the input and generated control signals (loudness,  centroid, pitch) only on non-silent (loudness $> -40dB$) frames. We measure loudness error in dBFS RMS and centroid error in semitones (st). We report the following pitch metrics: pitch error (st), chroma error (from the predicted pitch), and periodicity error, as estimated by torchcrepe \cite{morrison2022torchcrepe}. We only measure pitch and chroma error on \textit{voiced} frames, i.e.,~where the periodicity predicted by torchcrepe is greater than a threshold of 0.5 for both the vocal imitation and the model output.

\section{Experiments}
\subsection{Control signals}
First, we validate that Sketch2Sound can synthesize sounds from a reference vocal imitation and a text prompt while achieving a competitive audio quality, text adherence, and adherence to the vocal imitation. We fine-tune three models, each with a different set of control signals (loudness only, loudness+centroid, loudness+centroid+pitch), and compare the performance of these models, along with a text-only baseline, using the metrics discussed in Section \ref{sec:exp-design}. For all model variants, we use a fixed median filter window size of 10 at inference. 

\subsection{Sketch type ablation}
To observe the effect of our random median filtering as a way of creating sketchlike controls, we compare our approach to a no-filter baseline, as well as an alternative approach using low-pass filters instead of median filters to remove fine temporal detail from the controls during training. 
We train our model with random median filters with window sizes ranging between 1 and 25 control frames (1  frame = 25 ms). At inference, we use a fixed median filter size of 10. For our low-pass filter approach, we apply random low-pass filters ranging from 5Hz-20Hz and use a fixed cutoff (10Hz) at inference. 
Likewise, we compare each variant in this experiment in terms of audio quality, text and control adherence, following  Section \ref{sec:exp-design}.

\subsection{Inference-time control rates}
To verify whether our median filtering approach allows for an inference-time trade-off between text adherence and fine-temporal control, we observe the performance of our model trained with random median filters at different inference-time temporal resolutions. Specifically, we generate samples using  our model at different inference-time median filter sizes of $\{1, 5, 10, 15, 20, 25\}$ observe their performance on the metrics (Section \ref{sec:exp-design}). 

\section{Results and Discussion}

\begin{figure}[t]
\centerline{\includegraphics[width=\linewidth]{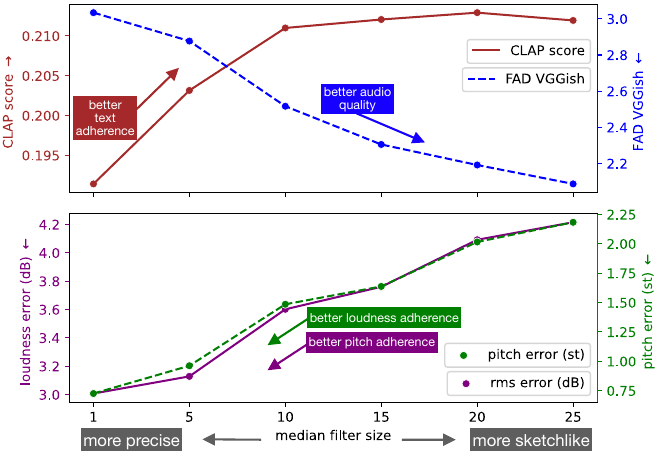}}
\captionsetup{labelformat=empty}
\caption[width=\columnwidth]{Fig. 2: At inference, larger median filters are more sketchlike and can lead to higher audio quality, while smaller filters are more precise and may lead to lower audio quality if the vocal imitations aren't precise enough, giving the sound artist a choice over this trade-off.\vspace{-12pt}
}
\label{fig:control-rates}
\end{figure}

\subsection{Control signals}
Table \ref{tab:combined_table}  validates Sketch2Sound's ability to synthesize sounds using control signals extracted from a vocal imitation + a text prompt while achieving comparable audio quality to a text-only baseline. We incrementally add each control (loudness, centroid, pitch), and observe each model's performance in terms of text adherence (CLAP score), control signal adherence, and audio quality (FAD).  

Conditioning a model on a time-varying control signal (e.g.,~loudness, centroid or pitch) significantly improves the adherence to that control signal compared to when the model is not conditioned on that control. 
Since loudness, centroid, and pitch are often correlated in natural sounds, incorporating a single conditioning (i.e.,~loudness) also slightly improves the control adherence for other control signals. 


In terms of text adherence and audio quality, introducing control signals produces a slight decrease in audio quality and text adherence. Empirically, we find that this difference is near negligible when compared to the text-only baseline in most cases. We also find that the quality of the generated audio can be a function of how well the user imitates the characteristics of the target sound: better-performed vocal imitations lead to higher-quality generated sound effects. 

\subsection{Sketch type ablation}

Table \ref{tab:combined_table} shows that our median filter method makes Sketch2Sound robust to generating high-quality sound effects from sketchlike vocal imitation control signals, improving the audio quality and text adherence over a no-filter baseline and a low-pass method. Notably, our median filter method is able to achieve a higher CLAP score (text adherence) and lower FAD (audio quality), while trading off the control adherence to vocal imitations. \textbf{This trade-off, where lowering the control adherence to vocal imitations improves the audio quality is more desirable than a strict adherence to the controls since most vocal imitations cannot perfectly mimic the fine temporal behavior of target sounds. Generating sounds that exactly follow the vocal imitation controls (i.e.~\textit{``no filtering''}) results in audio that does not sound like the text, but ``speechlike''.}



\subsection{Inference-time control rates}

Our control-rate filtering method lets a sound artist use different-size median filters at inference time, allowing users to choose the desired amount of temporal detail needed for a particular voice-to-sound example. 
The results in Figure 2 show that Sketch2Sound can be used with different control-rate resolutions at inference time by using median filters of different sizes. Smaller filters achieve higher control adherence, at the cost of a lower audio quality and text adherence. We hypothesize that the decrease in text adherence and audio quality is due to the mismatch between vocal control signals and the target control signals suitable for generating Foley sounds. 
However, this flexibility allows one to use smaller filters (i.e.,~higher temporal resolution) when the vocal imitations are well-performed, and larger filters (i.e.,~lower temporal resolution) when the vocal imitations are impossible to precisely imitatate with the human voice.   

\subsection{The semantics of control curves are implicitly modeled}
\vspace{-9pt}
\begin{figure}[hbtp]
\centerline{\includegraphics[width=\linewidth]{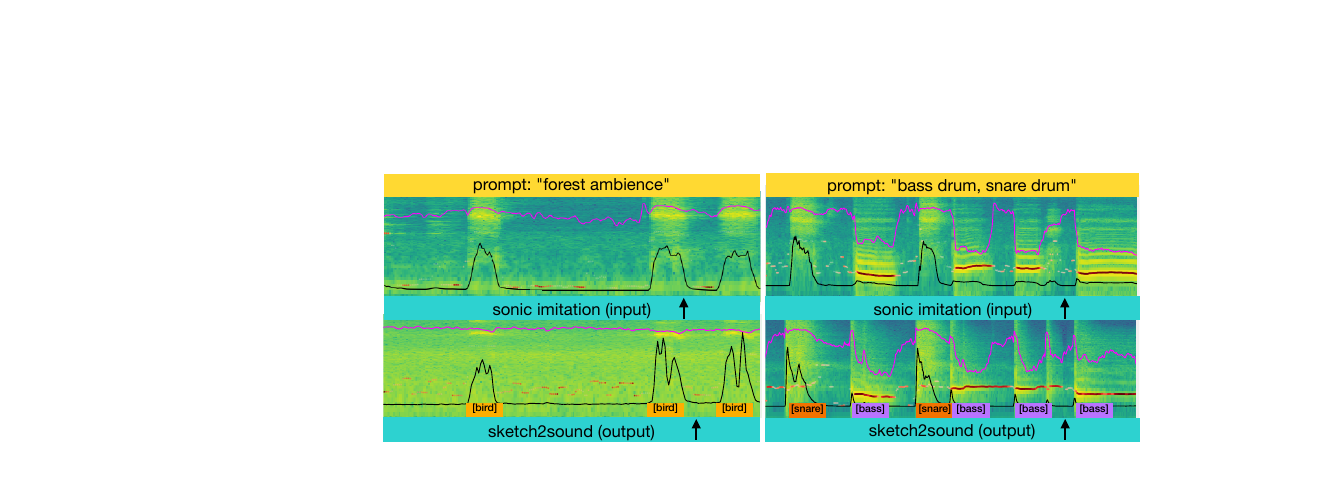}}
\captionsetup{labelformat=empty}
\caption[width=\columnwidth]{Fig. 3: (left) When prompted with ``forest ambience'', bursts of loudness in the controls become of birds without prompting the model to do so. (right) With ``bass drum, snare drum'', the model places snares in unpitched areas and bass drums in pitched areas.  }
\end{figure}

We find that the control signals can manipulate the semantics of the generated signals. 
For example, using the text prompt ``forest ambience'' with a sonic imitation containing random bursts of loudness in it, we can synthesize bird sounds into those loudness bursts (see Figure 3), without having to use the prompt ``birds'' at all. The model follows the correlations between bursts of loudness and the presence of birds in recordings of forest ambience in the training data, and so generates bird sounds when prompted with loudness bursts in a ``forest ambience'' prompt. 
Likewise, with the prompt ``snare drum, bass drum, drum beat'', performing a sonic imitation with a series of pitched (bass drum) and unpitched (snare drum) sounds will successfully apply bass drum sounds on pitched regions, and snare drum on unpitched regions. 
Both of these examples are available on our accompanying website. 
\subsection{Limitations}

We found that the centroid control tends to entangle the room tone of the input sonic imitation onto the generated audio. We believe this is because the room tone of the input audio is encoded by the centroid when no sound events are occurring in the input audio. A potential solution is to drop the controls when the signal is quiet, though this may fail with overcompressed inputs or loud room tones.

\section{Conclusion}
Sketch2Sound is a generative sound model capable of generating sounds from text prompts and time-varying controls: loudness, brightness (spectral centroid), and pitch. 
Sketch2Sound is capable of generating sounds from sonic imitations and sketchlike control curves. Sketch2Sound is a lightweight adapter method, requiring only 40k steps of fine-tuning and a single linear adapter layer per control signal.  
We showed that Sketch2Sound can create arbitrary sounds via sonic imitations with a flexible amount of temporal specificity while retaining the semantics of a text prompt, making it a controllable, gestural, and expressive tool for sound artists.    

\section*{Acknowledgments}
H.F.G. would like to thank Zachary Novack for many meaningful conversations on diffusion, generative audio models, statistics, and median filters. The authors thank Rithesh Kumar for his insight into conditional generative modeling and Adolfo Hernandez Santisteban for valuable discussions in interface design and interaction for sound design practices. 

\bibliographystyle{IEEEtran}  
\bibliography{references}  

\vspace{12pt}

\end{document}